\documentclass[12pt]{article}
\usepackage{amssymb}                         % -я¬-я¬-я¬-я¬-я¬-я¬-я¬-я¬-я¬-я¬-я¬-vя¬- я¬-я¬-я¬-я¬-я¬-я¬-v
\usepackage[T2A]{fontenc}
\usepackage{enumerate,amsmath,graphicx}
\usepackage{graphicx}
\usepackage{amssymb}

\newcommand{\R}{\mathbb{R}}

\newcommand{\Z}{\mathbb{Z}}

\newtheorem{algorithm}{Algorithm}
\newtheorem{theorem}{Theorem}
\newtheorem{lemma}{Lemma}
\newtheorem{corollary}{Corollary}

\begin{document}

\pagestyle{plain}
\title{Approximate Solution of Length-Bounded Maximum Multicommodity Flow
with Unit Edge-Lengths\thanks{This research is supported by the
Russian Science Foundation grant 15-11-10009.}}

\author{Pavel Borisovsky$^1$, Anton Eremeev$^1$,
Sergei Hrushev$^1$,\\
        Vadim Teplyakov$^2$,
        Mikhail Vorozhtsov$^2$}

\maketitle
\begin{center}
$\mbox{ }^1$ Sobolev Institute of Mathematics, Siberian Branch of
              Russian Academy of Sciences,
              4~Koptyug Ave., 630090 Novosibirsk, Russia\\
$\mbox{ }^2$ Yaliny Research and Development Center,
              Paveletskaya naberezhnaya~2 block~2, Moscow, 115114, Russia
\end{center}

\begin{abstract}
An improved fully polynomial-time approximation scheme and a
greedy heuristic for the fractional length-bounded maximum
multicommodity flow problem with unit edge-lengths are proposed.
Computational experiments are carried out on benchmark graphs and
on graphs that model software defined satellite networks to
compare the proposed algorithms and an exact linear programming
solver. The results of experiments demonstrate a trade-off between
the computing time and the precision of algorithms under
consideration.\\


\end{abstract}

%
% \usepackage{mathptmx}      % use Times fonts if available on your TeX system
%
% insert here the call for the packages your document requires
%\usepackage{latexsym}
% etc.
%
% please place your own definitions here and don't use \def but
% \newcommand{}{}
%
% Insert the name of "your journal" with
% \journalname{myjournal}
%

\section{Introduction}

Research issues in telecommunications often involve requirements
to quality of service such as bandwidth, delay, number of hops,
etc.~\cite{M03,TZ09}. These research issues may be
interpreted as combinatorial optimization problems of finding a
multicommodity flow through the network that satisfies some
quality of service constraints.

Multicommodity flow (MCF) problems are defined on a directed graph
$G=(V,E)$ with edge capacities $u:E\to \mathbb{R}$ and $k$
origin-to-destination pairs $(s_j,t_j),$ $j=1,\dots,k$. These
problems ask for a family of flows $f_j$ from $s_j$ to $t_j$, so
that some optimization criterion is maximized under the node flow
conservation constraints and the requirement that the sum of flows
on any edge does not exceed the capacity of the edge.

In particular, the {\em maximum multicommodity flow} problem ({\em
maximum MCF} for short) is an MCF problem where the total flow
needs to be maximized. Some authors assume that the maximum MCF
also includes an assumption that the value of each flow~$f_j$ is
limited by a finite demand~$d_j, \ j=1,\dots,k$ (see
e.g.~\cite{TZ09}).

A more general problem, called {\em fractional length-bounded
maximum multicommodity flow} ({\em fractional length-bounded
maximum MCF} for short), asks for computing the maximum MCF routed
along a set of paths whose length does not exceed a specific
bound. This problem was studied in~\cite{Baier03,B00} assuming
unbounded demands and in~\cite{CPR04}, assuming that finite
demands are given. In particular, it was shown in~\cite{Baier03}
that the fractional length-bounded maximum MCF is NP-complete,
while its special case where all edges have unit length is
polynomially solvable.

A modification of the fractional length-bounded maximum MCF with
an additional constraint that the flow on all edges must be
integer-valued is called integral length-bounded maximum MCF. The
results from~\cite{GVY97} show that even when there is no length
constraint at all, the edge capacities are equal to~1, and the
graph is outerplanar, this problem does not admit approximation
algorithms with constant approximation ratios, unless P=NP. A
result from~\cite{GKR99} implies that the integral length-bounded
maximum MCF can not be approximated with a performance ratio
$1/n^{0.5-\varepsilon}$ for any $\varepsilon>0$ in the special
case where all edges have equal length and all edge capacities are
equal to~1, assuming P$\ne$NP. Complexity and non-approximability
of this problem was also studied in~\cite{Baier03,Baier10}.

Fully polynomial time approximation schemes~(FPTAS) for fractional
length-bounded maximum MCF were proposed in~\cite{Baier03,CPR04}.
In~\cite{TZ09}, an FPTAS was shown to exist for a more general
{\em quality of service-aware MCF problem} (QoS-aware max-flow).
The present paper is aimed at fast approximate solution of a
special case of fractional length-bounded maximum MCF where all
edges have the length equal to one.

In Section~\ref{sec:routing_formulation}, we give a formal
statement of fractional length-bounded maximum MCF and describe
some basic properties of this problem. In
Section~\ref{sec:approx}, using the approach from~\cite{GK98} and
the improvement of~\cite{Fleis2000} we develop an FPTAS for a
special case of the problem where all edges have unit length. This
FPTAS has a smaller time complexity compared to the FPTAS
from~\cite{TZ09} which was developed for arbitrary edge lengths.
In Section~\ref{sec:greedy}, we propose a simple greedy heuristic
applicable to fractional length-bounded maximum MCF as well as
integral length-bounded maximum MCF (assuming unit edge lengths in
both cases). In Section~\ref{sec:exp}, we compare these algorithms
to each other and to IBM~CPLEX solver in computational
experiments. The last section contains conclusions.
% and further research questions.

\section{Problem Formulations and Basic Properties}
\label{sec:routing_formulation}

A flow in a digraph~$G=(V,E)$ from origin vertex~$s\in V$ to
destination vertex~$t \in V$ is a nonnegative function~$f:E\to
R_+$ such that for each node $v \in V, \ v\ne s, \ v\ne t$ holds
$$
\sum_{(v',v)\in E} f(v',v) - \sum_{(v,v')\in E} f(v,v') =0
$$
and
$$
|f|=\sum_{(s,v')\in E} f(s,v') - \sum_{(v',s)\in E} f(v',s)
 = \sum_{(v',t)\in E} f(v',t) - \sum_{(t,v')\in E} f(t,v')
$$
is the amount of flow sent from~$s$ to~$t$ in~$f$.
%On one hand,
%any flow~$f$ may be represented as a sum of at most a polynomial
%number of {\em path flows} $f^{P_1},\dots,f^{P_r}$ where each
%flow~$f_j$ is positive only on edges of path~$P_j, \ j=1,\dots,r.$
%The path flows $f^{P_1},\dots,f^{P_r}$ may be found
%by an augmenting path algorithm, e.g. from~\cite{Karz}.
%On the other hand,
If $P_1,\dots,P_r$ are paths from~$s$ to~$t$,
then a sum of path-flows along $P_1,\dots,P_r$ gives a network
flow from~$s$ to~$t$ again. Given that $f=\sum_{j=1}^r f^{P_j}$,
we will say that~$f$ {\em is routed along the set of
paths}~$P_1,\dots,P_r$.
% (there may be a number of ways of
%doing this).

We will assume that flow~$f_i$ of commodity~$i, i=1,\dots,k,$ has
an origin~$s_i\in V$ and a destination~$t_i \in V$. If
$f_1,f_2,\dots,f_k$ are flows of $k$ commodities, then
$F=(f_1,f_2,\dots,f_k)$ is called a {\em multicommodity flow}
in~$G$.

% DEFINE $\mathbb{R}^+$ and $\mathbb{R}_+$ !!!
An input instance of the maximum MCF consists of a directed
network ${G=(V,E)}$, where $|V|=n, \ |E|=m$, an edge capacity
function ${u:E \to \R_+}$ and a specification~$(s_i,t_i,d_i)\in
V\times V\times \R^+$ of commodity~$i$ for $i=1,\dots,k$. The
objective is to maximize $\sum_{i=1}^k |f_i|,$ so that the sum of
flows on any edge~$e\in E$ does not exceed~$u(e)$ and $|f_i|\le
d_i, \ i=1,\dots,k$.

The fractional length-bounded maximum MCF has the same input,
extended by an upper bound~$L\in \Z^+$ and the edge
lengths~$\tau(e),\ e\in E$ and asks for a maximum MCF where the
sum of flows on any~$e\in E$ does not exceed~$u(e)$, $|f_i|\le
d_i, \ i=1,\dots,k,$ and the flow of each commodity is routed
along a set of paths, where each path has a length at most~$L$. A
special case of this problem where all edges have a unit length
requires that the flow of each commodity is routed along a set of
paths at most~$L$ edges long. In what follows, we will denote
fractional length-bounded maximum MCF problem by LBMCF and its
special case where all edges have unit length will be denoted by
LBMCF1. Obviously, the maximum MCF may be considered a special
case of LBMCF1, assuming~$L=n$. W.l.o.g. we will assume that all
pairs~$(s_i,t_i)$ are unique, since otherwise the demands with
identical pairs~$(s_i,t_i)$ may be summed together in one demand.

LBMCF1 is polynomially equivalent to its special case with
unbounded demands. Indeed, given an LBMCF1 instance ${G=(V,E)}$,
${u:E \to \R_+}$, $(s_i,t_i,d_i),$ $i=1,\dots,k$ and~$L$, consider
a new instance of this problem with unbounded demands on a
network~$G'$, obtained from~$G$ as described below. Let
$T_v\subset V$ be the set of all vertices where at least one
commodity originating in~$v$ is consumed, i.e. $T_v:=\cup_{i:
s_i=v} \{t_i\}$. For each vertex~$v\in V$ with $|T_v|>0$:
\begin{itemize}
\item New $|T_v|$ vertices
are created and connected by arcs leading into vertex~$v$.

\item The new vertices are assigned to commodities
with destinations in $T_v$ by a one-to-one mapping. So we can
denote the new vertices connected to~$v$ as $v_i$ where~$i$ is
such that~$s_i=v$.

\item The capacities of edges leading from vertices~$v_i$ to $v$ are
set to~$d_i$.
\end{itemize}
In the new problem the demands are~$(v_i,t_i,+\infty),$
$i=1,\dots,k$ and the new upper bound is~$L'=L+1.$

%Obviously,
There is a bijection between the sets of feasible solutions of the
original instance and the new instance  and the objective function
values of the corresponding solutions are equal.
Now since the special case of LBMCF1 with unbounded demands is
solvable by LP methods in weakly polynomial time~\cite{Baier03},
the above reduction implies that LBMCF1 with finite demands is
also solvable in weakly polynomial time (the same follows from the
LP formulation of this
problem~(\ref{eqn:max_flow1})--(\ref{eqn:Balance1}) presented
below).
The question of solvability of LBMCF1 in strongly polynomial time
remains open. In the special case of maximum MCF, a strongly
polynomial algorithm is known~\cite{Tardos}. Nevertheless, even in
this special case existence of exact algorithms using the same
combinatorial techniques as the Ford-Fulkerson method for the
single-commodity flow is unlikely (see
e.g.~\cite{Schreiver_C},~\S~70.13).

In the special case of maximum MCF there is a well-known
formulation the problem in terms of the LP using edge flows (see
e.g.~\cite{Hu},~\S~11) with $O(k+m)$ constraints and $O(km)$
variables. Assuming that variables ${x_{i}(e)\ge 0}$ give the
amount of flow of commodity~$i$ over edge~$e$ the LP model is as
follows.

\begin{equation} \label{eqn:max_flow1}
\max \sum_{i=1}^k \ \sum_{e=(s_i,v) \in E} \  x_{i}(e),
\end{equation}

\begin{equation} \label{eqn:demand1}
\sum_{e=(s_i,v) \in E} x_{i}(e) \le d_i, \ \ i=1,\dots,k,
\end{equation}

\begin{equation} \label{eqn:bandwidth1}
\sum_{i=1}^k x_{i}(e) \le u(e), \ \ e\in E,
\end{equation}

\begin{equation} \label{eqn:Balance1}
 \sum_{e=(v',v) \in E} x_{i}(e)
 =
 \sum_{e=(v,v') \in E} x_{i}(e),
 \ \ i=1,\dots,k, \ \ v \in V\backslash \{s_i,t_i\},
\end{equation}

An LP formulation of LBMCF1, involving~${O(Lkn+m)}$ constraints
and $O(Lkm)$ variables, may be constructed using a multicommodity
flow in a supplementary time-expanded network~\cite{KS06}. The
node set~$V'$ contains a copy~$V_t$ of the node set~$V$ of
graph~$G$ for every discrete time step~$t, t=1,\dots,L$. For every
directed edge~$(v,w)\in E$ there is an edge in~$E'$ from
vertex~$v_t\in V_t$ in time layer~$t$ to vertex~$w_{t+1} \in
V_{t+1}$.
%connecting the
%corresponding vertices in the time-expanded network.
Besides that,~$E'$ contains edges $(v_t,v_{t+1})$ for all $v_t\in
V_t, \ t=1,\dots,L-1.$ A multicommodity flow is sought in this
time-expanded network under additional constraints which require
that for each $e=(v,w)\in E$ the sum of all flows traversing the
edges $(v_t,w_{t+1}), \ t=1,\dots,L-1$ is at most~$u(e).$ For
all~$i=1,\dots,k,$ the origin of commodity~$i$ is placed in the
copy~$s_{i1}$ of vertex~$s_i$ at level~1 and the destination is
placed in the copy~$t_{iL}$ of vertex~$t_i$ at level~$L$. The
resulting LP formulation is as follows

\begin{equation} \label{eqn:lev_max_flow1}
\max \sum_{i=1}^k \ \sum_{e'=(s_{i1},v_2) \in E'} \  x_{i}(e'),
\end{equation}

\begin{equation} \label{eqn:lev_demand1}
\sum_{e'=(s_{i1},v_2) \in E'} x_{i}(e') \le d_i, \ \ i=1,\dots,k,
\end{equation}

\begin{equation} \label{eqn:lev_bandwidth1}
\sum_{i=1}^k \sum_{e'=(v_t,w_{t+1})\in E'} x_{i}(e') \le u(e), \ \
e=(v,w)\in E,
\end{equation}

\begin{equation} \label{eqn:Balance1_}
 \sum_{e'=(v_{t-1},w_{t}) \in E'} x_{i}(e')
 =
 \sum_{e'=(w_{t},v_{t+1}) \in E'} x_{i}(e'),
\end{equation}
$$
 i=1,\dots,k, \ \ w_t \in V_t, \ \ t=2,\dots,L-1,
$$

\begin{equation} \label{eqn:Balance2}
 \sum_{e'=(v_{L-1},w_{L}) \in E'} x_{i}(e')
 = 0,
 \ \ i=1,\dots,k, \ \ w_L \in V_L \backslash \{t_{Li}\},
\end{equation}

where variables $x_{i}(e')\ge 0$ give the amount of flow of
commodity~$i$ over edge~$e'\in E'$.

The practice shows that large MCF problems require a long time and
a great amount of memory to solve using the exact LP methods
either in path flow-based or edge flow-based formulations. (see
e.g.~\cite{Schreiver_C},~\S~70.13). For this reason, it is
important to develop faster algorithms to solve MCF problems
approximately and LBMCF1 among them.

A feasible solution~$y$ to a maximization problem is called a {\em
$(1-\omega)$-approxi\-mate} if it satisfies the inequality $ f(y)
\ge (1-\omega) f^*,$ where $\omega>0$ and $f^*$ is the optimal
objective function value. An algorithm is called a {\em
$(1-\omega)$-approximation algorithm} if in a polynomially bounded
time it outputs a $(1-\omega)$-approximate solution given a
solvable problem instance. A family of $(1-\omega)$-approximation
algorithms parameterized by~$\omega>0$, such that the time
complexity of these algorithms is polynomially bounded in
$1/\omega$ and in the problem instance length is called {\em a
fully polynomial-time approximation scheme~(FPTAS)}.
% (see, e.g.~\cite{GJ}).

%In~\cite{CPR04,TZ09}, FPTASes were proposed for LBMCF problem and
%the next section presents a faster FPTAS for the case of LBMCF1
%problem.

\section{Fully Polynomial Time Approximation Scheme} \label{sec:approx}

\subsection{The Case of Unbounded Demands} \label{FPTAS1}

This subsection presents an FPTAS for LBMCF1 with unbounded
demands, which is developed analogously to the FPTAS for maximum
MCF with unbounded demands~\cite{Fleis2000}.
%nevertheless it is given here for the sake of completeness.

Let ${\mathcal P}_i$ denote the set of all paths from~$s_i$
to~$t_i$ in~$G$ and let ${\mathcal P}_i(L)$ denote the subset
of~${\mathcal P}_i$ which consists of paths at most~$L$ edges
long. Besides that, put ${\mathcal P}(L)=\cup_{i=1}^k {\mathcal
P}_i(L).$
%Finally,
%let~$X_i(P)\in R_+$ denote the amount of flow of commodity~$i$
%routed along path~$P\in {\mathcal P}_i$.
The main difference from the preceding
algorithms~\cite{GK98,Fleis2000} is that instead of~${\mathcal
P}_i,\ i=1,\dots,k$ here we use~${\mathcal P}_i(L), i=1,\dots,k$
and search for the shortest paths in~${\mathcal P}_i(L)$ by means
of a truncated version of Ford-Bellman algorithm.

LBMCF1 in the case of unbounded demands may be formulated as an LP
problem (denoted by~{\bf P}) with an exponential number of path
flow variables $x(P), \ P\in {\mathcal P}(L):$

\begin{equation} \label{eqn:crit1inf}
\max \sum_{P\in {\mathcal P}(L)} x(P),
\end{equation}

\begin{equation}\label{eqn:capasinf}
\sum_{P\in {\mathcal P}(L) : e \in P}  x(P) \le u(e), \ \ e\in E,
\end{equation}

\begin{equation}\label{eqn:posinf}
x(P) \ge 0, \ \ P\in {\mathcal P}(L).
\end{equation}

The dual problem with a polynomial number of
variables $y(e) \ge 0, \ \ e \in E$ is

\begin{equation}\label{eqn:dualcrit1}
\min \sum_{e \in E} u(e) y(e),
\end{equation}

\begin{equation}\label{eqn:packing_prime}
\sum_{e \in P} y(e) \ge 1, \ \ P \in {\mathcal P}(L),
\end{equation}
\begin{equation}\label{eqn:dualpos1_prime}
 y(e) \ge 0, \ \ e \in E,
\end{equation}

We first describe the general ideas of the
$(1-\omega)$-approximation algorithm for problem~{\bf P} according
to the framework of Garg and K\"{o}nemann~\cite{GK98} and after
that a faster version will be described in detail.

The algorithm proceeds by iterative improvements of primal
solutions to problem~{\bf P} simultaneously computing a sequence
of dual feasible solutions. The latter ones allow to estimate the
precision of the current primal solution and to find directions
for further improvement. It is convenient to compute a set of
parameters $\{\ell(e)\}_{e\in E}$, called {\em edge lengths}
instead of the current dual-feasible solution~$y$. The dual
feasible~$y$ may be reconstructed by scaling $y(e)=\ell(e)/\alpha,
\ e\in E$ where the factor~$\alpha$ is the length of a shortest
path in~${\mathcal P}(L)$.

The algorithm starts with length function~$\ell(e)=\delta$ for all
$e\in E$ using some $\delta>0$, and with a primal
solution~$x(P)=0, P\in {\mathcal P}(L)$. While there is a path
in~${\mathcal P}(L)$ of length less than~1, the algorithm selects
such a path and updates the primal and the dual variables as
follows. For the primal solution~$x$, the algorithm increases the
flow along path~$P$ by the minimum edge capacity in the path. Let
us denote this bottleneck capacity by~$u$.
%The primal solution is then updated by
%setting $x(P) = x(P)+u$.
The updated primal solution may be infeasible, so in order to
return~$x$ to feasible region, all of its components are scaled
down by an appropriate scalar. Now the dual variables are updated
so that the higher the congestion of an edge the greater
multiplier is given to its length:
$$
\ell(e) = \ell(e)\left(1 + \frac{\varepsilon u}{u(e)}\right), \
e\in P.
$$
In particular, the length of the bottleneck edge always increases
by a factor of~$(1 + \varepsilon)$. The lengths of edges not
on~$P$ remain unchanged.

%It will be shown that this algorithm has only a polynomial number
%of iterations, and each iteration increases~$x(P)$ for just one
%path~$P$, therefore the number of non-zero components in~$x$ is
%polynomially bounded.

In order to find out if there is a path~$P\in{\mathcal P}(L)$ of
length~$\ell(P)<1$ according to the current length function, it
suffices to compute a shortest path for each commodity, e.g. by
executing~$L$ iterations of Bellman-Ford algorithm, which would
take a total of~$O(kLm)$ time (see e.g. Theorem~2.3 in~\cite{NW}).
Let $\alpha:=\min_{P\in {\mathcal P}(L)} \ell(P)$ denote the
current shortest path length and let~$\hat{\alpha}$ be a lower
bound on~$\alpha$, which will be evaluated implicitly at each
iteration of the algorithm as described below.

Instead of looking through all origin-destination pairs of
commodities, seeking for a shortest path in~${\mathcal P}(L)$, we
implement the improvement of L.~Fleischer~\cite{Fleis2000} which
consists in using a path of length at most~$(1+\varepsilon)\alpha$
and spending less time to find such a path. To this end, we cycle
through all commodities, staying with one commodity until the
shortest origin-to-destination path for that commodity is
above~$(1+\varepsilon)\hat{\alpha}$. The initial lower bound
$\hat{\alpha}$ is set to~$\delta$. As long as there is some
path~$P\in{\mathcal P}(L)$ of length
$\ell(P)<\min\{1,(1+\varepsilon)\hat{\alpha}\}$, we augment the
flow along such~$P$. When such a path does not exist, it means
that either $\alpha\ge 1$ and it is time to terminate the
algorithm or $\alpha<1$ and $\alpha\ge (1+\varepsilon)
\hat{\alpha}$. In the latter case one can update the lower bound
by setting~$\hat{\alpha}:=(1+\varepsilon) \hat{\alpha}$. With such
updates, the lower bound will belong to the
set~$\{\delta(1+\varepsilon)^r\}_{r=0,1,2,\dots}$. Upon the
termination of the algorithm, $\hat{\alpha}\in [1,1+\varepsilon]$.
Since each time~$\hat{\alpha}$ is increased by a factor of
$(1+\varepsilon)$, the number of times that this happens is
$\lfloor \log_{1+\varepsilon}
\frac{1+\varepsilon}{\delta}\rfloor$, where $\lfloor \cdot\rfloor$
denotes the rounding down. This implies that the final value
of~$r$ is $r_{\max}=\lfloor \log_{1+\varepsilon}
\frac{1+\varepsilon}{\delta}\rfloor$. Assuming that every new
value of the lower bound~$\hat{\alpha}$ defines a new phase~$r$ of
the algorithm, we obtain an FPTAS represented by
Algorithm~\ref{alg:FPTAS} with the main loop over
phases~$1,\dots,r_{\max}$ as described below.

If there is a vertex~$v'\in T_v$ such that no path from~${\mathcal
P}_i(L)$ leads from~$v$ to~$v'$, then the corresponding demand can
not be served at all. All such pairs of vertices $v,v'$ may be
identified at the preprocessing stage and the corresponding flows
should be set to zero. W.l.o.g. we will assume that such pairs of
vertices do not exist.
%For any $v'\in T'_v$, we will denote
%by~$i(v,v')$ the number of commodity~$i$, such that $s_i=v$ and
%$t_i=v'$ (this number is unique according to assumption made in
%Section~\ref{sec:routing_formulation}).

The detailed outline of the algorithm follows the
$(1-\omega)$-approximation algorithm for maximum
MCF~\cite{Fleis2000} with minor modifications.

In what follows we assume that $\ell(P)$ denotes the sum of
lengths of all edges comprising a path~$P,$ and
%$S$ is the set of
%all vertices where at least one commodity originates, i.e.
 $S:=\cup_{i=1}^k \{s_i\}$. By a shortest path we mean a path
 from~$\mathcal{P}(L)$ with minimal length~$\ell(P)$.\pagebreak

\begin{algorithm}{\bf $(1-\omega)$-approximation algorithm for
LBMCF1 with unbounded demands} \label{alg:FPTAS}
\end{algorithm}

%\vspace{-1em}
{\bf Initialization}\\
%Let $\varepsilon =\frac{3-\omega-\sqrt{(3-\omega)^2-4\omega}}{2}$.
Choose $\varepsilon, \delta$
%as in Theorem~\ref{theor:FPTAS} and
%Corollary~\ref{cor:FPTAS} below
and assign
 $r_{max}:=\lfloor \log_{1+\varepsilon}
 \frac{1+\varepsilon}{\delta}\rfloor$.\\
Assign $l(e) := \delta$ for all $e \in E$.\\

{\bf The main loop}\\
$\mbox{ }$ {\bf For all} $r=1,...,r_{max}$ {\bf do}\\
$\mbox{ }$  \  \  \   {\bf For all} $v \in S$ {\bf do}\\
$\mbox{ }$ \  \  \   \  \  \ Find $|T_v|$
shortest paths~$P(v') \in\mathcal{P}(L)$ from~$v$ to all $v'\in T_v.$\\
$\mbox{ }$ \  \  \   \  \  \   Choose $v^*=\mbox{arg}\min\limits_{v' \in T_v} \ell(P(v'))$. Let $P=P(v^*)$.\\
$\mbox{ }$ \  \  \   \  \  \   {\bf While} $\ell(P) < \min\{1,
\delta(1 +
\varepsilon)^r\}$ {\bf do}\\
$\mbox{ }$ \  \  \   \  \  \   \  \  \   \  \  \   Let $u:=\min_{e \in P} u(e)$.\\
$\mbox{ }$ \  \  \   \  \  \   \  \  \   \  \  \   For all $e \in
P$ assign
$\ell(e):=\ell(e)\big(1+\frac{\varepsilon u}{u(e)}\big)$.\\
$\mbox{ }$ \  \  \   \  \  \   \  \  \   \  \  \   Augment the
path-flow $x(P):=x(P)+u/\log_{1+\varepsilon} \frac{1+\varepsilon}{\delta}$. \\
$\mbox{ }$ \  \  \   \  \  \   \  \  \   \  \  \   Find $|T_v|$
shortest paths in~$P(v') \in\mathcal{P}(L)$ from~$v$ to all $v'\in T_v.$\\
$\mbox{ }$ \  \  \   \  \  \   \  \  \   \  \  \ Choose
$v^*=\mbox{arg}\min\limits_{v' \in T_v} \ell(P(v')).$ Let $P=P(v^*)$.\\
$\mbox{ }$ \  \  \   \  \  \   {\bf End while}\\
$\mbox{ }$ \  \  \   {\bf End for-loop over $v$}\\
$\mbox{ }$ {\bf End for-loop over $r$}.\\

In the above algorithm, the length-bounded shortest paths from~$v$
are found by performing only~$L$ iterations of Bellman-Ford
algorithm.
% (see e.g.~\cite{CLRC01}).\\
%
Correctness of Algorithm~\ref{alg:FPTAS} and its time complexity
are established analogously to those of $(1-\omega)$-approximation
algorithm for maximum MCF~\cite{Fleis2000}, therefore the
following Theorem~\ref{theor:FPTAS} is given without a proof here.
(The proof may be found in Appendix.)

\begin{theorem} \label{theor:FPTAS}
(i) Algorithm~\ref{alg:FPTAS} completes after
   $O\left( m \log_{1+\varepsilon} \frac{1+\varepsilon}{\delta} \right)$
   augmentations.

(ii) Given
$\delta=\frac{1+\varepsilon}{\sqrt[\varepsilon]{(1+\varepsilon)L}},$
the feasible solution to LBMCF1 with unbounded demands found by
Algorithm~\ref{alg:FPTAS} has a total flow value~$g \ge \frac{f^*
(1-\varepsilon)^2}{1+\varepsilon}. $
\end{theorem}

Suppose a value $\omega>0$ is given. Then choosing $\varepsilon
=\frac{3-\omega-\sqrt{(3-\omega)^2-4\omega}}{2}$ and
$\delta=\frac{1+\varepsilon}{\sqrt[\varepsilon]{(1+\varepsilon)L}}$,
by Theorem~\ref{theor:FPTAS}, part~(ii) we conclude that the
obtained flow has a value at least $(1-\omega)f^*$.

The length-bounded shortest paths from vertex~$v$ are computed
once in the beginning of each iteration in the loop over~$v\in S$,
additionally it is computed after each augmentation. In total
there are at most~$|S|\cdot r_{max}=O(\varepsilon^{-2} n \log(L))$
applications of the truncated Bellman-Ford algorithm in
Algorithm~\ref{alg:FPTAS} that do not lead to augmentations.
Besides that, Theorem~\ref{theor:FPTAS}, part~(i) implies that the
truncated Bellman-Ford algorithm  is applied at most~$O(m
\log_{1+\varepsilon}
\frac{1+\varepsilon}{\delta})=O(\varepsilon^{-2} m \log(L))$ times
with augmentations. Thus the total runtime of
Algorithm~\ref{alg:FPTAS} is~$O(\varepsilon^{-2} m^2 L \log(L))$
and we have the following

\begin{corollary} \label{cor:FPTAS}
Given~$\omega\in (0,1),$ Algorithm~\ref{alg:FPTAS} with
$\varepsilon =\frac{3-\omega-\sqrt{(3-\omega)^2-4\omega}}{2}$ and
$\delta=\frac{1+\varepsilon}{\sqrt[\varepsilon]{(1+\varepsilon)L}}$
computes a $(1-\omega)$-approximate solution to LBMCF1 with
unbounded demands in $O(\omega^{-2} m^2 L \log L)$ time.
\end{corollary}

%This corollary gives a lower time complexity estimate compared to
%that of FPTAS~\cite{TZ09} for the more general problem LBMCF in
%the case of unbounded demands, which has the upper bound
%${O(n\omega^{-2} m^2 \log m \cdot (\log \log n +1/\omega))}$ on
%the runtime.

The idea of adapting the approach from~\cite{Fleis2000} to the
length-bounded MCF has already been discussed by
Baier~\cite{Baier03} for the case with general edge lengths. It
was already noted in~\cite{Baier03} that the resource-constrained
shortest path problem, which arises as a subproblem in the general
case, is NP-hard and can only be solved approximately. In the
present paper, we exploit the fact that in the unit edge length
case, the subproblem becomes efficiently solvable.

\subsection{The Case of Finite Demands}

\label{subsec:UQoS_FPTAS}

The reduction described in Section~\ref{sec:routing_formulation}
allows to obtain $(1-\omega)$-approximate solutions to LBMCF1 by
applying Algorithm~\ref{alg:FPTAS} to the transformed instance
which consists of network~$G'$ with $n'=n+k$ vertices and $m'=m+k$
edges, the specification of commodities~$(v_i,t_i,+\infty), \
i=1,\dots,k$ and the length bound~$L'=L+1$. We can save some time
by using the structure of~$G'$ where one can find the shortest
paths from all new vertices~$v_i$ attached to a vertex~$v, \ v\in
S,$ by a single execution of truncated Bellman-Ford algorithm,
starting with vertex~$v$. Then in total there are at
most~$O(\varepsilon^{-2} n \log(L))$ calls of the Bellman-Ford
algorithm that do not lead to an augmentation and there are at
most~$O(\varepsilon^{-2} (m+k) \log(L))$ of calls to the
Bellman-Ford algorithm following the augmentations and we have
%the following corollary from Theorem~\ref{theor:FPTAS}.

\begin{corollary} \label{cor:FPTAS_UQoS}
A $(1-\omega)$-approximate solution to LBMCF1 may be computed in
$O(\omega^{-2} (m+k)m L \log L)$ time.
\end{corollary}

This corollary gives a lower time complexity bound compared to the
time bound $ {O(\omega^{-2} (m+k)mn \log (m+k) \cdot (\log \log n
+1/\omega))} $ of the FPTAS developed in~\cite{TZ09}. However the
FPTAS from~\cite{TZ09} is applicable to LBMCF where the length
function has a more general form.

%Note that the technique applied here for the case of unit
%edge-lengths not seem to be applicable to general length
%functions, as the respective constrained shortest path problem is
%NP-hard.

\section{Greedy Heuristic}\label{sec:greedy}

As an alternative to the guaranteed approximation algorithm from
Subsection~\ref{subsec:UQoS_FPTAS}, we propose a simple Greedy
heuristic, based on augmenting paths, for LBMCF1. In what follows,
by shortest path we mean a path with the minimal number of edges
in~$G$.

The main idea of Greedy heuristic consists in iterative assignment
of augmenting path-flows to the most distant origin-destination
pairs (but not more than~$L$ edges apart). In each iteration of
Greedy, the shortest paths are found for all origin-destination
pais and a maximum possible flow is routed along the path of
maximal length among the paths with at most~$L$ edges. After that
we decrease the edge capacities along the path by the value of its
flow, delete all edges where the remaining capacity has turned
to~0 and proceed to the next iteration.
%If the full demand can
%not be delivered along the chosen shortest path, then the
%heuristic will attempt to route the undelivered amount of
%commodity along the shortest path(s) in the network with trimmed
%set of edges in the subsequent iterations.
The algorithm terminates when a set~$\mathcal S$ of shortest paths
from~${\mathcal P}(L)$, connecting the unsatisfied
origin-destination pairs in the current network, becomes empty.
Here we assume that~${\mathcal P}(L)$ is the set of paths at
most~$L$ edges long {\em in the current network.}

%Greedy heuristic  has
%the following outline.
%In what follows, $m(P)$ denotes the
%number of edges in path~$P$.

\begin{algorithm}{\bf Greedy Heuristic for LBMCF1} \label{alg:greedy_mod}
\end{algorithm}
%\vspace{-1em}

\noindent {\bf Initialization of set ${\mathcal S}:$}\\
$\mbox{ }$ Compute a shortest path $P_{s_i t_i}\in {\mathcal
P}(L)$ from~$s_i$~to~$t_i$\\
$\mbox{ }$ for all $i=1,\dots,k$, for which such paths exist.\\
$\mbox{ }$ Denote the set of computed paths by ${\mathcal S}$.\\
{\bf The main loop:}\\
$\mbox{ }$  {\bf While} $|{\mathcal S}|>0$ {\bf do}\\
$\mbox{ }$    \  \  \ Choose a longest path $P$ in
${\mathcal S}$, put ${\mathcal S}:={\mathcal S} \backslash P$.\\
$\mbox{ }$  \  \  \ Let $s_i$ and $t_i$ be the first and the last vertices in path~$P$.\\
$\mbox{ }$  \  \  \ $u_{\min}:=\min\{u(e): e \in P\}$,  $x(P):=\min\{u_{\min},d_i\}$.\\
$\mbox{ }$  \  \  \ $d_i:=d_i-x(P)$. \\
$\mbox{ }$  \  \  \ {\bf For} all~$e\in P$ {\bf do} $u(e):=u(e)-x(P).$ \\
$\mbox{ }$  \  \  \ {\bf If} there are
edges~$e\in P$, such that $u(e)=0$ {\bf then}\\
$\mbox{ }$  \  \  \  \  \  \  \ Delete all edges~$e\in P$, such that $u(e)=0$, from~$G$. \\
$\mbox{ }$  \  \  \  \  \  \  \ Build a new set ${\mathcal S}$ by
computing shortest paths $P_{s_i t_i}\in {\mathcal P}(L)$\\
$\mbox{ }$  \  \  \  \  \  \  \ from~$s_i$~to~$t_i$ for all
$i=1,\dots,k$, for which such paths exist.
\\
$\mbox{ }$  \  \  \  {\bf End if}.\\
$\mbox{ }$  {\bf End while.}\\

Clearly, the collection of path-flows with~$x(P)>0$ found by
Greedy constitutes a feasible solution.
%to LBMCF1 problem
Besides that, since the set~$\mathcal S$ of length-bounded
shortest paths is computed at most~$m$ times, the time complexity
of Greedy heuristic is~$O(m^2\log n)$ if the Dijkstra algorithm
with heaps is used.
%(see
%e.g.~\cite{CLRC01}).
If the truncated Bellman-Ford algorithm is used, then the time
complexity of Greedy is~$O(m^2 L)$.

Greedy can easily be converted to compute approximate solutions
for fractional length-bounded maximum MCF if instead of the number
of edges in a path one takes the path length in terms of edge
lengths~$\tau(e)$. Note that Greedy outputs a feasible solution
with an integer-valued flow on all edges if an instance is
feasible and all demands and edge capacities are integer-valued.

\section{Computational Experiments} \label{sec:exp}

This section describes the computational experiments in solving
LBMCF1 by $(1-\omega)$-approximation algorithm from
Subsection~\ref{subsec:UQoS_FPTAS}, by Greedy heuristic from
Section~\ref{sec:greedy} and by the LP-solver CPLEX~11 in dual
simplex mode, using the LP formulation based on the time-expanded
networks. For the large instances, where this formulation required
too much time and memory we used the LP formulation of Maximum MCF
from~\cite{Hu} to find an upper bound for the optimum.
%In all
%experiments $(1-\omega)$-approximation algorithm was run with
%$\omega=0.2$ setting.
All experiments were carried out on Xeon~X5675, 3.07~GHz,
96~Gb~RAM, 12~cores.

\subsection{Implementation Details}

The $(1-\omega)$-approximation algorithm for LBMCF1 was
implemented with a minor improvement~\cite{Fleis2000} which allows
to terminate the algorithm when the best-found primal solution
%(under appropriate scaling)
and the best-found dual solution are within the required
approximation ratio. Whenever the amount of some commodity~$i$,
routed by the current iteration, achieved the corresponding
demand~$d_i$, this commodity was excluded from consideration in
subsequent iterations.

It is easy to see that the most time-consuming part of the Greedy
algorithm is the procedure building the shortest paths tree. We
found that this part of Greedy is so compute-intensive that CPU
cache misses percent may slow down the execution significantly, so
we reduced the memory footprint of the all-pairs shortest paths
procedure by implementing compact data structures both for the
input graph and for the temporary data. Also we used a specially
designed version of the Dijkstra's algorithm that may be called in
parallel for each root vertex ${v\in V}$. With these
optimizations, we reduced the final time-footprint of the Greedy
up to a factor 0.1 of its first na\"\i ve implementation.

\subsection{Problem Instances}

The networks for testing instances were obtained using a
modification of
%single-commodity maximum-flow problem
generator RMFGEN~\cite{gold}, besides that 7 instances with
real-life structure were constructed.
RMFGEN produces a given number of two-dimensional grids with arcs
connecting a random permutation of nodes in $b$ adjacent planar
grids of size~$a\times a$. Arcs of the grids have capacities~3600,
while the capacities of arcs connecting the grids are chosen
uniformly at random from~1 to~100. Origins and destinations are
randomly chosen.
%The parameters $a$ and $b$ are encoded in the names of the
%instances which have a form ``rmfgen-$a$-$b$''.

The demands were assigned in two alternative modes. In mode~I, the
demands were scaled as in~\cite{gold} so that there is a feasible
flow with the maximum arc congestion~$\lambda$, where $\lambda$ is
set to~0.6 or~1. In mode~II, each randomly chosen
origin-destination pair~$s_i, t_i\in V^2$ was assigned a random
factor~$\gamma_i$ from~1 to~100 and the total multicommodity flow
was maximized under constraint that the origin-destination
flows~$|f_i|$ are proportional to factors~$\gamma_i$. Finally we
assign the demand values~$d_i=1.2\times |f_i|$ to test the
algorithms in situation where the demands exceed the network
capacities.

The instances with real-life structure model prospective Software
Defined Satellite Networks~(SDSN) (see e.g.~\cite{TZYFW14}) that
provide world-wide telecommunication services. We suppose that the
network consists of 135~satellites on low Earth orbit~(LEO),
40~ground stations (gates to Internet) and a Network Operations
Control Center~(NOCC). The packet routes for each
origin-to-destination pair~$(s_i,t_i),$ are computed at NOCC in
real time and each node (satellite or ground station) regularly
receives the updated routs for all packets that originate in this
node. Each packet sent from $s_i$ to $t_i$ contains some content
data and a path of the packet route from $s_i$ to $t_i.$ An upper
bound~$L$ on the number of edges in packet paths is imposed due to
a natural technical limitation on the number of bits reserved for
encoding a packet route. Short packet paths also tend to have low
transmission delay. For simplicity we assume that each problem
instance describes the system in a single time-frame and all
demands for the time-frame are known in advance.
%and there is enough time to distribute the
%packet routing data computed at NOCC to all source nodes by the
%beginning of the time-frame.

The graphs modelling the SDSN were constructed with different
trade-off between the model accuracy and the size of~$G$. Networks
of these instances contain vertices of low degree (from~2 to~7),
corresponding to satellites and dummy nodes, and vertices with
high degree (near to~$n/3$), corresponding to Internet gates at
ground stations. The demands for commodities were generated so as
to model the global telecommunication flows. We assumed that the
number of active users in each square unit of the Earth surface is
proportional to population on the unit. The origin and the
destination of each call is chosen at random among active users.
All active users are assigned to the nearest satellite or ground
station.

\begin{table}[h]
\caption{Parameters of Instances With Real-Life Structure}
\label{tabl:Problems}
 \centering
\begin{tabular}{||c|c|c|c|c|c|c|c||}
\hline
{ Instance }    &1& 2& 3 & 4 & 5& 6 &7\\
\hline
$n$ &543&272& 272 & 197 & 197& 135 & 318\\
$m$ &19474 &  1292 & 1292& 992 & 992& 750 & 4652\\
$k$ &12373&12373 & 12373& 12373 & 12373 &743 & 245481\\
  \hline
\end{tabular}
\end{table}

\subsection{Experimental Results}

The values of relative errors of solutions found by FPTAS and by
Greedy heuristic were estimated a-posteriori in terms of upper
bound~$\omega':=(UB-f_{\rm appr})/UB$, where~$f_{\rm appr}$ is the
value of objective function found by the algorithm and $UB$~is the
optimum in the LP model based on the time-expanded network (on
grid graphs) or an upper bound on the optimum in maximum MCF LP
formulation (on the instances with real-life structure).
%The value of~$UB$ was found by CPLEX, using LP formulation of
%Maximum MCF problem~(\ref{eqn:max_flow1}) -- (\ref{eqn:Balance1}).

\paragraph {Grid Graphs.}

The attained approximations~$\omega'$ and CPU times (in seconds)
for the grid graphs with~$a=6$ and $k=15$ are given in
Table~\ref{tabl:Bandwidth}. Here demands are generated in mode~I.
One can see that the FPTAS occupies the position between Greedy
heuristic and CPLEX solver both in terms of the precision and the
running time, except for the two smallest instances where Greedy
was able to find optimal solutions. The parallel version of Greedy
using 12 cores is clearly the fastest one, achieving speed-ups of
about 4.5~times compared to the serial version. On small instances
this speed-up vanishes due to communication cost.

\begin{figure}
\begin{center}
\vspace{2em}
\includegraphics[width=12cm]{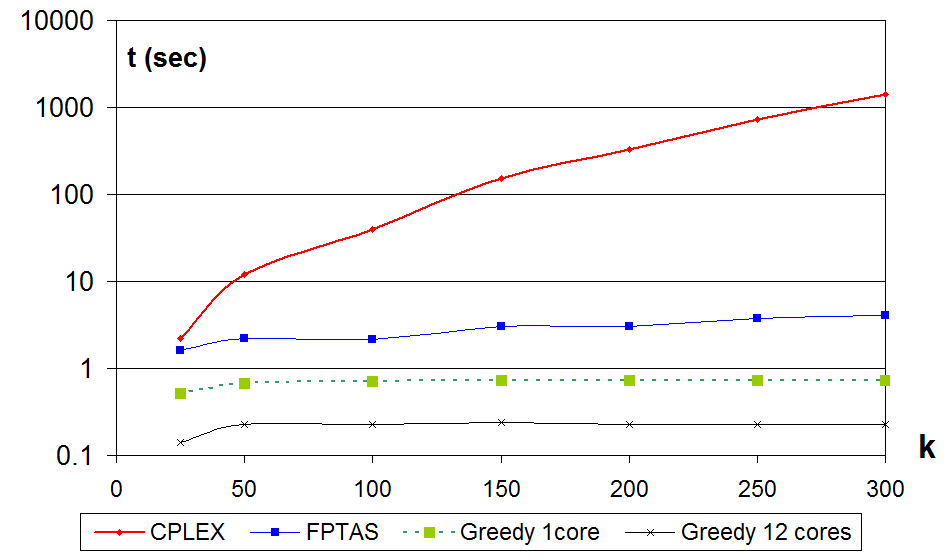}\\
\caption{CPU time as function of~$k$ on grid graph, $a=6$, $b=8$.
Demands generated in mode~II.\label{fig4}}
\end{center}
\end{figure}

The growth of CPU times with further increase of~$k$ is displayed
in Fig~\ref{fig4}. Here we use the largest grid graph with $a=6$,
$b=8$. The demands are generated in mode~II. Both approximate
algorithms have run-time upper bounds independent of~$k$ (see
Sections~\ref{sec:approx} and~\ref{sec:greedy}), which
 agrees with Fig~\ref{fig4} where the corresponding curves
are nearly horizontal. The size of LP formulation based on the
time-expanded network depends significantly on~$k$ and this is
supported by Fig~\ref{fig4}.

\begin{table}[h]
\caption{A posteriori estimated approximation~$\omega'$ and CPU
times on grid graphs. Here $w=0.2, L=9$. } \label{tabl:Bandwidth}
 \centering
 \begin{tabular}{||c|c|c||c c|c||c|c||c||}
\hline
\multicolumn{3}{||c||}{ RMFGEN} &\multicolumn{3}{c||}{Greedy}& \multicolumn{2}{c||}{FPTAS,}& CPLEX\\
\hline
$b$ & $\lambda$&$|E|$&\multicolumn{2}{c|}{CPU time}&$\omega'$&CPU time&$\omega'$& CPU time \\
\hline
                         &&& 1    & 12 &            & 1    && 1  \\
                         &&& core & cores &      & core      && core \\
\hline \hline
2  & $0.6$ & 276 &        0.01 & 0.01 & {\bf 0} & 0.07 & 0.01 & 0.15\\
4  & $0.6$ & 588 &        0.05 & 0.02 & {\bf 0} & 0.23 & 0.01 & 0.25 \\
6  & $0.6$ & 900 &        0.09 & 0.02 & 0.07    & 0.53 & {\bf 0.02} & 0.64\\
8  & $0.6$ & 1212 &       0.14 & 0.03 & 0.29    & 0.74  & {\bf 0.02} & 1.06\\
2  & $1$ & 276 &       0.01 & 0.01  & 0.04    & 0.08 & {\bf 0.04} & 0.15 \\
4  & $1$ & 588 &       0.06 & 0.02    & 0.1     & 0.3 & {\bf 0.05} & 0.61 \\
6  & $1$ & 900 &       0.14 & 0.04    & 0.07    & 0.57 & {\bf 0.01} & 0.44\\
8  & $1$ & 1212 &      0.18 & 0.04    & 0.27    & 0.79 &
{\bf 0.02} & 0.59\\
  \hline
\end{tabular}
\end{table}

\paragraph {Instances With Real-Life Structure.}

Instances 1-7 have a greater range of graph sizes and a much
greater number of commodities compared to the instances with grid
graphs (see Tables~\ref{tabl:Problems} and \ref{tabl:Bandwidth}).

The LP model based on the time-expanded network required
prohibitive amount of time and memory. Therefore in the case of
Instances 1-7 we could compute only an upper bound~$UB$ using the
LP CPLEX solver. Still, in the case of Instance~1, CPLEX was
unable to find an upper bound due to lack of memory and $UB$ was
set to the total value of demands.

Table~\ref{tabl:Bandwidth1} shows the a~posteriori pessimistic
estimates of approximations~$\omega'$ attained by the algorithms
and the corresponding CPU times (in seconds). Here FPTAS always
has a greater precision than Greedy and the latter one is up
to~$10^3$ times faster even in the sequential version. Note that
%Greedy in all cases found approximate solutions with
%$\omega$ at most~$0.2,$ as we required from the FPTAS.

\begin{table}
\caption{A posteriori estimated approximation~$\omega'$ and CPU
times on instances with real-life structure. Here $w=0.2, L=9$.}
\label{tabl:Bandwidth1}
 \centering
\begin{tabular}{||c||c c|c||c|c||c||}
\hline
{ Inst-              } &\multicolumn{3}{c||}{Greedy}& \multicolumn{2}{c||}{FPTAS }& CPLEX\\
{ ance}       & \multicolumn{3}{c||}{heuristic}& \multicolumn{2}{c||}{ }& computing $UB$\\
\hline
         & \multicolumn{2}{c|}{CPU time} & $\omega'$ & CPU time & $\omega'$& CPU time \\
         & 1 core & 12 cores &  &  & &  \\

\hline
1       & 1.63 & 0.4  & 0.108    & 1689.6& {\bf 0.004} & -- \\
2       & 0.51 & 0.16 &0.072    & 157.6 & {\bf 0.008} & 7 971 \\
3       & 0.28 & 0.1  & 0.094    & 160.4 & {\bf 0.01} & 6 947\\
4       & 0.15 & 0.09 &  0.081    & 80.6 & {\bf 0.012} & 1 852\\
5       & 0.11 & 0.06 & 0.094    & 83.6  & {\bf 0.012} & 1 872\\
6       & 0.28 & 0.1  & 0.117    & 3.8   & {\bf 0.12} & 345\\
7       & 0.26 & 0.11 & 0.109 & 389.5    & {\bf 0.006} & 144 456 \\
  \hline
\end{tabular}
\end{table}

In order to evaluate the algorithms on a variety of different
real-life instances with similar structure we generated 300
versions of graph~$G$ that model SDSN analogously to Instance~7
but in different time-frames. Different satellite positions in
these time-frames lead to different links between satellites and
between satellites and ground stations. The bandwidth of the links
varies as well. The specification of demands remained unchanged.
Application of FPTAS and Greedy to these instances  in the
single-core version  and the same parameters as in
Table~\ref{tabl:Bandwidth1} gave the results shown in
Tables~\ref{tabl:300CPU} and~\ref{tabl:300omega}. The average and
maximum CPU times and $\omega'$ estimates are close to those
reported in Table~\ref{tabl:Bandwidth1} for Instance~7, which
implies that both algorithms have a stable behavior on the input
data that we considered.

\begin{table}
\caption{CPU times (sec.) on 300 instances with real-life
structure, $L=9$} \label{tabl:300CPU}
 \centering
\begin{tabular}{||c||c||c|c|c||}
\hline
{      } &{Greedy}& \multicolumn{3}{|c||}{FPTAS}\\
\hline
{      } &{      }& {$\omega=0.4$ }& {$\omega=0.2$ }& {$\omega=0.1$ }\\
\hline
Average & 0.3      & 164.8            & 402.0    & 857.6\\
\hline
Maximum & 0.38     & 177.6            & 444.1    & 888.7 \\
  \hline
\end{tabular}
\end{table}

\begin{table}
\caption{A posteriori estimated approximation~$\omega'$ on 300
instances with real-life structure, $L=9$} \label{tabl:300omega}
 \centering
\begin{tabular}{||c||c||c|c|c||}
\hline
{      } &{Greedy}& \multicolumn{3}{|c||}{FPTAS}\\
\hline
{      } &{      }& {$\omega=0.4$ }& {$\omega=0.2$ }& {$\omega=0.1$ }\\
\hline
Average & 0.061      & 0.006           & 0.004    & 0.003\\
\hline
Maximum & 0.148      & 0.01            & 0.006    & 0.004\\
  \hline
\end{tabular}
\end{table}

\section*{Conclusions}

The proposed FPTAS has a lower time complexity bound compared to
the previously known algorithms designed for a problem with the
length functions of more general form.

The FPTAS and Greedy heuristic proposed in this paper are
significantly faster than the CPLEX LP solver, especially on the
instances with large networks and great numbers of demands. The
FPTAS is more accurate but requires more CPU time than Greedy
which may be a decisive factor in practical applications.
Implementation of the FPTAS (hopefully) may be improved using line
search for updating the current primal and dual solutions as
proposed in~\cite{Albrecht2001} and by the means of parallel
computations.

The exact LP-model discussed in
Section~\ref{sec:routing_formulation} is based on a Kirchoff-type
formulation in an extended graph. The alternative column
generation approach has no polynomial time-bound but is often more
efficient in practice. Further research might include a comparison
of the algorithms presented here to the column generation method,
assuming that columns are the length-bounded paths.

\section*{Appendix}

Correctness of Algorithm~1 and its time complexity are established
analogously to those of $(1-\omega)$-approximation algorithm for
maximum MCF~\cite{Fleis2000}. Before the proof of Theorem~1 we
formulate and prove two lemmas. The reason why these proofs were
omited in the paper is because they almost literally repeat the
corresponding proofs in~\cite{Fleis2000}.

 \setcounter{theorem}{0}

\begin{lemma}
  Algorithm~1 terminates after
   $O\left( m \log_{1+\varepsilon} \frac{1+\varepsilon}{\delta} \right)$
   augmentations.
\end{lemma}

{\bf Proof.} At start, $\ell(e) = \delta$ for all edges $e$. The
last time the length of an edge is updated, it is on a path of
length less than one, and it is increased by at most a factor
of~$1+\varepsilon$. Thus the final length of any edge is at
most~$1+\varepsilon$. Since every augmentation increases the
length of some edge by a factor of at least~$1+\varepsilon$, the
number of possible augmentations is at most~$m\log_{1+\varepsilon}
\frac{1+\varepsilon}{\delta}$. $\Box$

\begin{lemma}
  The flow
  $
  f(e)=\sum_{P\in {\mathcal P}(L): e\in P} x(P), \ e\in E
  $
  obtained by
  Algorithm~1 satisfies the constraints
  ${f(e) \le u(e)}$ for all $e \in E$.
\end{lemma}
{\bf Proof.} Every time the total flow on an edge~$e$ increases by
a fraction $0 < a_j\le 1$ of~$u(e)/\log_{1+\varepsilon}
 \frac{1+\varepsilon}{\delta}$, its length is multiplied
by~$1+a_j\varepsilon$. Since $1+a \varepsilon \ge
(1+\varepsilon)^a$ for all $0\le a \le 1,$ we have $\Pi_j (1 +
a_j\varepsilon) \ge (1 + \varepsilon)^{\sum_j a_j}$, when $0 \le
a_j \le 1$ for all~$j$. Thus, every time the flow on an edge
increases by its capacity divided by~$\log_{1+\varepsilon}
 \frac{1+\varepsilon}{\delta}$, the length of the edge
increases by a factor of at least $1 + \varepsilon$. Initially
$\ell(e) = \delta$ and at the end $\ell(e) < 1 + \varepsilon$, so
the total flow on edge~$e$ cannot exceed $u(e)$. $\Box$

\begin{theorem} %\label{theor:FPTAS}
(i) Algorithm~1 completes after
   $O\left( m \log_{1+\varepsilon} \frac{1+\varepsilon}{\delta} \right)$
   augmentations.

(ii) Given
$\delta=\frac{1+\varepsilon}{\sqrt[\varepsilon]{(1+\varepsilon)L}},$
the feasible solution to LBMCF1 problem with unbounded demands
found by Algorithm~1 has a total flow value~$g \ge \frac{f^*
(1-\varepsilon)^2}{1+\varepsilon}. $
\end{theorem}

{\bf Proof.} Part~(i) follows from Lemma~1.

Now consider part~(ii). Let $\ell_j$ denote the length function
after~$j$-th augmentation in Algorithm~1 and let $\alpha(\ell)$
denote the length of a shortest path in~${\mathcal P}(L)$ w.r.t. a
length function~$\ell$. Given a length function~$\ell$, define
$D(\ell):= \sum_e \ell(e)u(e),$ and let $D(j) := D(\ell_j).$ Then
$D(j)$ is the dual objective function value corresponding
to~$\ell_j$ and $\beta := \min_{\ell} D(\ell)/\alpha(\ell)$ is the
optimal dual objective value. Let $g_j$ be the primal objective
function value after $j$-th augmentation and let $P$ be the
augmenting path. Denote $\sigma:=\log_{1+\varepsilon}
\frac{1+\varepsilon}{\delta}$. Then for each $j\ge 1$,
$$
D(j)=\sum_e \ell_j(e) u(e)=\sum_e \ell_{j-1}(e) u(e) +
 \varepsilon \sum_{e\in P} \ell_{j-1}(e) u
 $$
 $$
 \le
 D(j-1) + \varepsilon (g_j-g_{j-1})(1+\varepsilon)\sigma
  \alpha(\ell_{j-1}),
$$
which implies that
\begin{equation} \label{eqn:2_2}
D(j) \le D(0) + \varepsilon (1+\varepsilon)\sigma\sum_{j'=1}^j
(g_{j'}-g_{j'-1}) \alpha(\ell_{j'-1}).
\end{equation}
Consider the length function $\ell_j-\ell_0$. Note that
$D(\ell_j-\ell_0)=D(j)-D(0)$. For any path used by the algorithm,
the length of the path using~$\ell_j$ versus $\ell_j-\ell_0$
differs by at most~$\delta L$. Since this holds for the shortest
path using length function $\ell_j-\ell_0$, we have
$\alpha(\ell_j-\ell_0) \ge \alpha(\ell_j)-\delta L.$ Hence
$$
\beta \le \frac{D(\ell_j-\ell_0)}{\alpha(\ell_j-\ell_0)} \le
\frac{D(j)-D(0)}{\alpha(\ell_j)-\delta L}.
$$
Using the bound on $D(j)-D(0)$ from equation~(\ref{eqn:2_2}), we
obtain
$$
\alpha(\ell_j) \le \delta L +
\frac{\varepsilon(1+\varepsilon)\sigma}{\beta} \sum_{j'=1}^j
(g_{j'}-g_{j'-1}) \alpha(\ell_{j'-1}).
$$
Observe that, for fixed~$j$, this right hand side is a
non-decreasing function on
$\alpha(\ell_0),\dots,\alpha(\ell_{j-1}).$ So, for any sequence of
upper bounds $\alpha'_j, \ j=1,\dots,$ on $\alpha(\ell_j), \
j=1,\dots,$ we have
$$
\alpha(\ell_j) \le \alpha'_j=\alpha'_{j-1}
(1+\varepsilon(1+\varepsilon)\sigma(g_j-g_{j-1})/\beta)\le
 \alpha'_{j-1} e^{\varepsilon(1+\varepsilon)\sigma(g_j-g_{j-1})/\beta},
$$
where the last inequality uses the fact that $1 + a \le e^a$ for
$a\ge 0$. We can use a valid upper bound $\alpha'_0 = \delta L$,
which implies
$$
\alpha(\ell_j) \le \delta L e^{\varepsilon (1+\varepsilon)\sigma
g_j/\beta}.
$$
By the stopping condition, after the last augmentation (let it be
the augmentation number~$t$) we have
$$
1\le \alpha(\ell_t)\le \delta L e^{\varepsilon
(1+\varepsilon)\sigma g_t/\beta}
$$
and hence
$$
\frac{g_t}{\beta} \ge \frac{\ln(\delta
L)^{-1}}{\varepsilon(1+\varepsilon)\sigma}=
 \frac{\ln(1+\varepsilon) \ln(L\delta)^{-1}}{\varepsilon(1+\varepsilon)
\ln\frac{1+\varepsilon}{\delta}}.
$$
Recalling that
$\delta=\frac{1+\varepsilon}{\sqrt[\varepsilon]{(1+\varepsilon)L}}$
we obtain
$$
\frac{g_t}{\beta} \ge \frac{(1-\varepsilon)
\ln(1+\varepsilon)}{\varepsilon(1+\varepsilon)}\ge
 \frac{(1-\varepsilon)
(\varepsilon -\varepsilon^2/2)}{\varepsilon(1+\varepsilon)}\ge
 \frac{(1-\varepsilon)^2}{1+\varepsilon}.
$$

 $\Box$



\begin{thebibliography}{}
% etc

\bibitem{Albrecht2001} Albrecht, Ch.: Global routing by new approximation algorithms
for multicommodity flow. IEEE Transactions on Computer-Aided
Design of Integrated Circuits and Systems. 20(5) 622 -- 632 (2001)

\bibitem{Baier03} Baier, G.: Flows with Path Restrictions.
Ph.D. Dissertation, TU Berlin, Berlin (2003)

\bibitem{B00}  Ben-Ameur, W.: Constrained-length connectivity and survivable
networks,  Networks 36(1) 170--33 (2000)

\bibitem{Baier10}
Baier, G., Erlebach, T., Hall, A., K\"{o}hler, E. Kolman, P.,
Pangr\'{a}c, O., Schilling, H. and Skutella, M.: Length-bounded
cuts and flows. ACM Trans. Algorithms, 7(1), 4:1--4:27 (2010)

%\bibitem{Bellman} Bellman, R.: On a Routing Problem. Quarterly of Applied
%Mathematics. 16 (1), 87--90 (1958)

%\bibitem{Bley03} Bley, A.: On the complexity of vertex-disjoint
%length-restricted path problems. Comput. Complex. 12 (3), 131--149
%(2003)

\bibitem{CPR04} Chaudhuri, K., Papadimitriou, C., Rao, S.:
Optimum routing with quality of service constraints. Unpublished
manuscript (2004).

%\bibitem{CLRC01}
%Cormen, T.H., Leiserson, C.E., Rivest, R.L., and Stein,~C.: {\em
%Introduction to Algorithms,} 2nd edition, MIT Press, 2001.

\bibitem{Fleis2000} Fleischer L.K.: Approximating fractional multicommodity flow independent of the
number of commodities, SIAM J.Disc.Math., 13, 505 -- 520 (2000)

%\bibitem{GJ} {Garey, M.R. and Johnson, D.S.}: Computers and intractability.
%A guide to the theory of NP-completeness. W.H. Freeman and
%Company, San Francisco (1979)

\bibitem{GK98} Garg, N., K\"{o}nemann, J.: Faster and simpler algorithms
for multicommodity flow and other fractional packing problems. In:
Proc. 39th IEEE Symposium on Foundations of Computer Science,
FOCS'98, pp. 300--309. IEEE CS Press (1998)

\bibitem{gold} Goldberg,~A.V., Oldham,~A.D., Plotkin,~S., Stein,~C.:
An implementation of an approximation algorithm for minimum-cost
multicommodity flows. In: Proc. of 6-th Integer Programming and
Combinatorial Optimization, LNCS vol. 1412, pp. 338-352, Berlin,
Springer (1998)

%\bibitem{GLS81} M. Gr\"{o}tschel, L. Lov\'{a}sz, and A. Schrijver.
%The ellipsoid method and its consequences in combinatorial
%optimization. Combinatorica, 1:169--197, 1981.


\bibitem{GVY97} Garg N., Vazirani V., Yannakakis M.
Primal-dual approximation algorithms  for integral flow and
multicut in trees, Algorithmica, 18, 3--20 (1997)

\bibitem{GKR99} Guruswami, V., Khanna, S.,  Rajaraman, R.,
Shepherd, B., and
Yannakakis, M.: Near-optimal hardness results and approximation
algorithms for edge-disjoint paths and related problems. In
Proceedings of the 31st Annual ACM Symposium on Theory of
Computing, pages 19--28 (1999)

\bibitem{Hu} Hu, T.C.: Integer Programming and Network Flows.
Reading, MA, Addison-Wesley Publishing Company (1970)

%\bibitem{HU} Hu, T.C.: Multi-commodity network flows. J. ORSA,
%1963, 11(3), 344-360.

%\bibitem{Karz} Karzanov, A.V.: Determining a maximal flow in a network
%by the method of preflows, Soviet Math. Dokl., 15, No.2, 434-437
%(1974)

\bibitem{KS06} Kolman, P., Scheideler, C.: Improved bounds for the
unsplittable flow problem. J. Algor., 61 (1), 20--44 (2006)

%\bibitem{MN97} Malashenko, Yu. E., Novikova, N. M.:
%Superconcurrent distribution of flows in multicommodity networks.
%Diskr. analiz i issled. oper., Ser. 2., No. 2, 34--54 (1997).

\bibitem{NW} Nemhauser, G.L. and Wolsey, L.A.:
Integer and Combinatorial Optimization. Wiley-Interscience. New
York, NY (1988)

\bibitem{M03} Van Mieghem, P., Kuipers, F.A., Korkmaz,~T., Krunz,~M., Curado,~M., Monteiro,~E., Masip-Bruin,~X., Sole-Pareta,~J.,
Sanchez-Lopez,~S.: Quality of service routing. In: Quality of
Future Internet Services. Lecture Notes in Computer Science, vol.
2856, pp. 80--117. Springer, Berlin (2003)

%\bibitem{PR94} Papaefthymiou M. and  Rodrigue J. Implementing parallel
%shortest paths algorithms, DIMACS Series in Discrete Mathematics
%and Theoretical Computer Science, pp. 59-68, 1994

%\bibitem{Radzik2001} Radzik, T.: Experimental study of a solution method for multicommodity flow
%problems. Journal of Experimental Algorithmics, 6, 79--102 (2001)

%\bibitem{Schr} Schrijver, A. Theory of Linear and Integer
%Programming. John Wiley \& Sons, Chichester, 1986.

\bibitem{Schreiver_C} Schrijver, A. Combinatorial Optimization.
Polyhedra and Efficiency. Vol.~C. Springer, 2003.

\bibitem{TZYFW14} Tang, Z., Zhao, B., Yu, W., Feng, Z and Wu, C.
Software defined satellite networks: Benefits and challenges. In:
Proc. of Computing, Communications and IT Applications Conference
(ComComAp), pp. 127-132. IEEE (2014)

\bibitem{Tardos} Tardos, E.: A strongly polynomial algorithm to solve combinatorial linear
programs.  Operations Research, 34 (2), 250-256 (1986)

%\bibitem{TS} Tarjan D., Skadron K., Micikevicius P. The Art of Performance Tuning for CUDA
%and Manycore Architectures.
%http://www.cs.virginia.edu/~skadron/Papers/cuda\_tuning\_bof\_sc09\_final.pdf

\bibitem{TZ09} Tsaggouris, G. and Zaroliagis, C.:
Multiobjective optimization: Improved FPTAS for shortest paths and
non-linear objectives with applications. Theory of Computing
Systems, 45 (1) 162--186 (2009)

%\bibitem{LinWang} Wang I-L. Shortest Paths and Multicommodity Network
%Flows. PhD dissertation. School of Industrial and Systems
%Engineering Georgia Institute of Technology, April~2003, 227~p.

%\bibitem{MPSP05} Wang I-L., Ellis L.J., Joel S.S.
%A multiple pairs shortest path algorithm. Transportation Science.
%Vol. 39, No. 4, November 2005, pp. 465-476.

\end{thebibliography}
\end{document}